\documentstyle[11pt,paspconf]{article}
\def\etal{{\it et al.}} 
\begin{document}
 
\title{Internal Kinematics of Dwarf Spheroidal Galaxies}
\author{Edward W.~Olszewski}
\affil{Steward Observatory, University of Arizona, Tucson, AZ, 85721-0065}
 
\begin{abstract}
We discuss the quality of kinematic data in dwarf spheroidal
galaxies, the current interpretations of those data, and prospects
for actually deriving the mass profile in these galaxies. We then
discuss stellar populations constraints on some of the models that
attempt to explain the kinematics and formation of the dSph's.
\end{abstract}

\section{Introduction}
It has now been fifteen years since Marc Aaronson (1983) used the MMT echelle
spectrograph to show that the CH stars in the Draco dwarf spheroidal (dSph)
had velocities inconsistent
with being drawn from a galaxy with luminosity M$_V$$=$$-$8.5 and with M/L$=$2, which
is the expectation for globular-cluster-aged population.
Instrumentation and detectors have improved markedly in that time, allowing
superb velocities to be obtained for stars in all of the dwarf spheroidals
orbiting the Milky Way.
The 
reader can find excellent reviews in Mateo (1994, 1998a,b), Da Costa (1998),
and Pryor (1994). 

\section{The Velocities Themselves, Present and Future}

In this section I will address the quality of the observational data.
The conclusions will be that the central
velocity dispersions of these galaxies are quite secure, and that the
properties of the 
stellar populations being measured do not corrupt
the line-of-sight velocities.

The typical star measured is a K giant. Fifteen years ago this was
not the case. Legitimate questions were raised about atmospheric
motions and a large binary population among the CH stars 
systematically inflating the measured velocity dispersion from the
true one. K giants are rather well-behaved stars; for instance,
Hatzes \& Cochran (1996, 1993) find that pulsations in
K giants occur at the 10--20 meter s$^{-1}$ level.
Samples of 30--300 stars with individual velocity errors of 2--3 km s$^{-1}$
are now available for all of the Milky Way dSph's.

The Ursa Minor and Draco galaxies, being the longest studied, have
multi-epoch velocities for more than 50 stars. Table 3 of Olszewski
\etal\ (1996a) shows that 41 stars have time baselines of 5 or more years,
with seven monitored for 10 or more years. 
Olszewski \etal\ (1996a) and Hargreaves \etal\ (1996)
have both shown that binary orbital motions 
do not inflate the measured dispersions
above the true dispersions. Comparisons of the velocities in Ursa
Minor measured by Armandroff \etal\ (1995) and Olszewski \etal\ (1995) and
Hargreaves \etal\ (1994) show that agreement is excellent, 
though with some low-level disagreements, finally
negating the complaint of Lynden-Bell \etal\ (1983) that velocities did not
agree within their quoted errors. 

Ultimately one would like line-of-sight velocity distributions,
which contain far more information than reducing all velocities
to a single datum, the velocity dispersion. In the case of
the Ursa Minor dwarf, which has one of the most extreme derived
dark matter densities, perhaps as high as 1 M$_\odot$ pc$^{-3}$
(Pryor \& Kormendy 1990), if we had all of the telescope
time we needed, how well could we do? First, Ursa Minor is a highly
resolved system, so long-slit spectra are out of the question.
Second, to B$\sim$22, there are 4.0 stars arcmin$^{-2}$ (Irwin \&
Hatzidimitriou 1995, hereafter IH). To a radius
of 1 core radius, which contains half of the stars in this system (UMi
is very different from a globular cluster in this respect), there
are 31 stars to V=18, 60 additional to V=19, 60 additional to V=19.5,
and 100 additional to V=20 (Cudworth \etal\ 1986). 
Since UMi is nearby, (m$-$M)$=$19.1, and metal poor, the fainter giants
and the subgiants are quite blue and 
weak lined, further adding to the observational
difficulties.
Samples of several hundred stars are going to be difficult to obtain.
The Fornax and Sgr dSph's are the only systems in which such
large samples will be available. Each has its own problems: Fornax has
a low heliocentric velocity, making samples unadulterated by Galactic
foreground stars hard to come by; Sgr is projected on
a very crowded part of the Milky Way.

\section {Tides}

Given the published velocity dispersions and King-model fits to
the projected density profiles, the dSph's are dark-matter dominated in
all cases (Mateo \etal\ 1993 and the reviews listed above). 
If the assumption that
these systems are in equilibrium is true, then large central
dark matter densities are needed and we have examples of
the smallest systems known to contain dark matter. If the systems
are out of equilibrium, then the large derived masses may be illusory.

Grillmair (1998) and Johnston (1998) 
discussed the tidal stripping
of stars from stellar systems. Grillmair
shows that globulars have lost most of their stars to 
tidal stripping, yet the derived M/L for globulars is ``normal.''
Simulations by Piatek \& Pryor (1995) and by Oh \etal\ (1995) show
that the central properties of dSph are little affected until the
object is unrecognizable as a dSph. Finally, we have good evidence
from its shape 
that Sgr is strongly affected by the Milky Way, yet its central velocity dispersion is,
remarkably, the same as that of the much more distant (25 kpc versus 140 kpc)
Fornax system,
which is nearest to Sgr in luminosity (Ibata \etal\ 1997, Mateo \etal\ 1991). 
In fact, Ibata \etal\
(1997) argue that while Sgr is being stripped, the fact that it has survived
at all means that it contains substantial dark matter.

Grillmair (1998), Piatek \&
Pryor (1995), and Oh \etal\ (1995) detail the evolution of a
globular cluster or a dwarf spheroidal when acted upon, gently or
strongly, by the Milky Way.
Grillmair and Piatek \& Pryor find that there are ``signature plumes''
from perigalactic passages. 
While these plumes are easily seen in the model, since they are
of low surface brightness and since we do not know the true shapes or
orientations of the dSph's, the plumes are not easily
seen in the density profiles. Piatek \& Pryor stress that the clearest
observed signature would be ordered motions along the two plumes. A
stellar sample would thus show a change in mean velocity along the major
axis (see Fig 8 of Piatek \& Pryor). 
Such a change is not observed in any dSph, except perhaps Sgr.

Kuhn \& Miller (1989) claimed that 
resonances between internal pulsational timescales of a dSph and its
orbital timescale about the Milky Way can produce departures from
virial equilibrium and also produce large velocity dispersions.
Pryor (1996) has raised a series of objections to this model, and
Sellwood \& Pryor (1998) have been unable to confirm this resonant
pumping.

Klessen \& Kroupa (1998, hereafter KK) and Kroupa (1997) have taken the point of view that
there are circumstances in which no-dark-matter tidal tails
can look remarkably like dSph's.
KK argue that if the orbit is
of eccentricity $>$0.5, then, when you look along the orbit, the debris
might look like a dSph. Furthermore, the ordered motions
described above would mask themselves as an increase in the
velocity dispersion, since all of the ordered motions would be
along the line of sight. These models are testable in
two ways: first, the true orbits can be derived from absolute
proper motions. Second, since one is observing them along the direction
of the tidal debris, one might expect that the color-magnitude diagram
would show that the dSph's all have large line of sight depths.
KK's Figure 18 shows predictions for the horizontal branch structure.
We give constraints on this model in the final section.

\section{Extra-Tidal Stars?}
A departure
of the dSph stellar density profile from that of the best-fitting King
model does not automatically imply that extratidal stars have been
detected. The long relaxation times in dSph's mean that
the velocity distribution is not necessarily Maxwellian. Given that the
velocity dispersion profiles discussed below imply that mass does not follow
light, at least two of King's (1962) assumptions
are violated. IH's claim that there
are extratidal stars is thus premature, as are conclusions drawn from
such a claim (Moore 1996, Burkert 1997).
In other studies, 
Gould \etal\ (1992) used deep multicolor photometry to find a few likely main-sequence stars
at the distance of the Sextans dwarf.
These stars
are 100 arcmin from the center of Sextans, which at the time was thought
to be beyond the tidal radius of Sextans. However,
the most recent star counts (IH) have shown
the tidal radius to be 160$\pm$50 arcmin. Kuhn \etal\ (1996) have
claimed that 25\% of the stars ``in'' Carina are outside its tidal
radius. UMi would be a good confirming object, since it is less
contaminated by foreground stars.
Of course, as discussed above, significant mass loss does not imply
that the measured velocity dispersions have been inflated.

\section{Masses and Mass Profiles}

The velocity dispersion profiles of $\sim$100 stars in Draco, $\sim$100
stars in UMi, and $\sim$200 stars in Fornax
tend to be rather flat, rather than to fall with increasing radius as 
a King model predicts.
Models in which mass follows light are tightly constrained and
seem to be ruled out by these data. There are two other reasons
for preferring models in which mass does not follow light.
First,
profiles in which mass does not follow light will provide the lowest
central dark matter densities, which already seem rather high.
Second, halos of dIrr and dwarf spiral galaxies
are substantially bigger than the optical extents of dSph galaxies.
Once we allow mass to be decoupled from light, it becomes much harder
to pin down the mass profile (Merritt 1987). 

The total masses of Draco and UMi can first be examined in a model-free way.
Figure 1 shows the minimum global M/L needed to bind each star
(Pryor \etal\ 1998). Applying single-component King models to
these galaxies, Pryor \& Kormendy (1990) calculate that the central-densities
are $\rho_{min}\sim 1.0 $ M$_\odot$ pc$^{-3}$ for Dra and UMi.
The velocity dispersion profile discussed above is a poor fit to
the predictions of the single-component models.
The luminous central densities are $\sim$0.03 M$_\odot$ pc$^{-3}$.
Minimum masses derived from the virial theorem are $\sim$2.7$\times$10$^7$
M$_\odot$, making M/L$\ge$10. 
The new, larger set of velocities now available gives $\rho_{min}=$0.2
M$_\odot$ pc$^{-3}$ for Draco and Umi (Pryor \etal\ 1998), using
2-component models described in Pryor \& Kormendy (1990); 
achieving such low central densities is only
possible with extended dark matter, thus these models have dark matter
by definition.

\begin{figure}[htb]
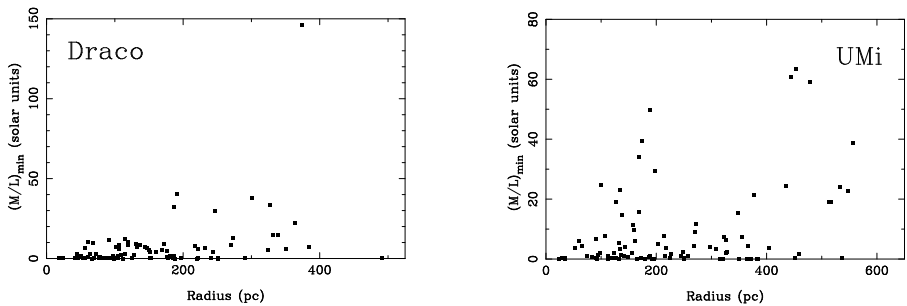

\plotfiddle{OLSZEWSKI.fig1_left.ps}{1.0 truein}{270}{23}{23} {-190}{100}
\plotfiddle{OLSZEWSKI.fig1_right.ps}{1.0 truein}{270}{23}{23} {0}{185}
\vspace{-0.75truein}
\caption{The minimum total mass, and thus M/L, 
needed to bind each star to the respective galaxy.
One or two extreme-velocity stars whose membership is open to question
have been left out of each plot. }
\end{figure}

Hans-Walter Rix has graciously run some models of Draco in which the input
is the luminous matter profile and the velocity dispersion at three 
radii. Orbits are calculated for three different potentials, Keplerian,
logarithmic (flat rotation curve), and harmonic (constant density),
and populated such that the luminous density profile and velocity dispersion
profile are matched (Figure 2). The h$_4$ parameter is a measure of
departure from a Gaussian distribution, with negative h$_4$ having more
stars in the velocity tails than does a Gaussian, and positive h$_4$ having
more stars at the systemic velocity and fewer in the wings.
The fourth row of panels shows
the relative contribution of the tangential and radial dispersion
profiles to the total dispersion, which will perhaps be
a measurable constraint when internal proper motions are available
for these systems (SIM and GAIA satellites).

Although the three
h$_4$ profiles are different, with our current measuring errors (Armandroff
\etal\ 1995), h$_4$
is constrained to $\pm$0.06 for the global set of 90 stars, and to
$\pm$0.13-0.15 for two bins of 45 stars. 
Doubling the sample and lowering the measuring errors
to 2--3 km s$^{-1}$ for every star will lower the errors in the h$_4$'s
to a level that may allow these three models to be distinguished.
\begin{figure}[t]
\plotfiddle{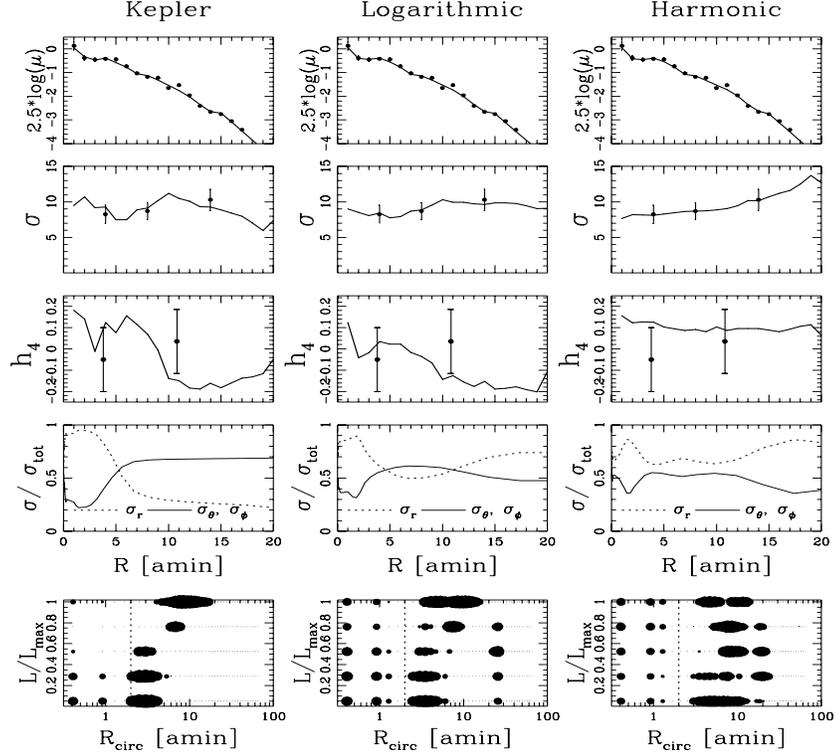}{4.0 in}{0}{60}{50} {-175}{-55}
\vspace{-0.3 truein}
\caption{Numerical modelling of Draco by H.-W.~Rix using Schwarzschild's
method (see Rix \etal\ 1997). Surface brightness profile of Draco taken from 
the literature (top
panels); Derived velocity dispersion profile overlaid on the data (second
row); h$_4$ statistic, the degree of departure from a Gaussian profile
(third row); contribution of radial and tangential dispersions to
the total velocity dispersion (fourth row); orbital weights(fifth row).
Estimates of the observed h$_4$ in a 2-bin Draco sample are superposed
on the h$_4$ plots.
}
\end{figure}

\section{Stellar-populations Constraints to Models}

KK's models require a substantial orbital eccentricity.
Schweitzer \etal\ (1995, 1998) have measured tangential velocities for
UMi and Sculptor. Kyle Cudworth has graciously allowed me to quote their
results: for UMi, v$_r$$=$$-$86$\pm$1, v$_t$$=$190$\pm$28; for Scl,
v$_r$$=$74$\pm$2, v$_t$$=$210$\pm$125. These results are in the Galactic
rest frame. Cudworth estimates ratios of apogalacticon distance to perigalacticon
distance of $\sim$2 ($e =$ 0.3). KK require large eccentricity and 
quote $e >$0.5, or r(apo)/r(peri)$>$3,
for their models to give large velocity dispersion. The more radial the
orbit the better their model succeeds.

The velocity dispersions are not the only issue.
Because such models are viewed along the orbit, there
will be a nonnegligible line-of-sight depth.
The magnitude dispersion for 84 blue horizontal branch stars in Ursa Minor
(Cudworth \etal\ 1986) is 0.09 mag. 
But 33 variables in M15 (Bingham \etal\ 1984) give a dispersion of
0.15 mag, and 35 variables in M3 (Sandage 1981) give a dispersion
of 0.07 mag. The line of sight depth of UMi
as measured by the horizontal branch is consistent with its projected
size as measured from star counts. 

The intriguing alignments of outer halo clusters and dSph's on the sky
(see Lynden-Bell \& Lynden-Bell 1995 and Majewski 1994 for the latest
papers on this subject) can be tested both by measuring orbits and by
stellar populations arguments. 
The best stream in the opinion of Lynden-Bell and Lynden-Bell is
LMC-SMC-UMi-Dra (with Sculptor and Carina as possible
members). Umi is approximately 180$^\circ$ away in the sky from the Magellanic
Clouds, so it must have been stripped long ago.

Ursa Minor and Draco are both as old as Galactic globulars (Olszewski \&
Aaronson 1985, Grillmair \etal\ 1998), and are
very metal poor with [Fe/H]$\sim$$-$2
(Suntzeff \etal\ 1984, Lehnert \etal\ 1992, Shetrone 1998). UMi also has
a blue horizontal-branch (HB) morphology (Cudworth \etal\ 1986).

The LMC has a ubiquitous red HB, with few, if any, BHB stars (see the
references in the 
review of Olszewski \etal\ 1996b, and Geha \etal\ 1998 for examples).
While the BHB of an old population is masked by the main-sequence of a younger one, there are other reasons to think that this old, metal-poor
population is very small. First,
there are remarkably few metal poor giant stars in a ``halo''
field defined by proper motions and complete spectroscopy (Olszewski
1993). This 0.25 deg$^2$ field, approximately 8 kpc from
the center of the LMC, formed one giant star, or 3\% of its mass
as revealed by red giants, when it was as metal poor as the dSphs. Finally, the pulsational
properties of the RR Lyraes imply evolution from a system with
a red HB (Alcock \etal\ 1996). The mean abundance of the RR Lyraes is
$-$1.3 to $-$1.8, again more metal rich than UMi or Dra.

There are few BHB stars in the SMC as well (Gardiner \& Hatzidimitriou
1992). The mean age and abundance of the ``halo'' field near NGC 121 is 8 Gyr
and [Fe/H]$=$$-$1.6, respectively (Suntzeff \etal\ 1986). About 20\%
of this ``halo'' is of the right abundance to make a Draco or UMi.

While the abundance arguments taken alone cannot rule out a scenario
in which 
UMi and Dra are LMC or SMC tidal fragments, the age-metallicity relations
for the Magellanic Clouds demand that this interaction happened ``in
the beginning.'' First, it then seems more reasonable to argue that all of
these galaxies were simply individual chunks of a protogalactic clump
coming together to make the Milky Way. 
Second, since UMi is at the antipode
of this LMC orbit, presumably
the whole orbit would be populated. Since we argue here that UMi must split off
at perigalacticon passage number 1, where are the results of the
remaining many such passages? Why was the LMC or SMC fragile enough to
be tidally disrupted on passage number one, but strong enough to
have survived a Hubble time? Is such a scenario possible? The age-metallicity
relations for the Magellanic Clouds, the properties of a ``halo'' field
in the LMC, and the properties of the LMC RR Lyraes seem to be daunting
foes of traditional tidal models.

Lin \& Murray (1994) and Lin (1996) advocate models in which dSph's shared
the Magellanic Cloud star-formation history until they were tidally removed. At tidal
splitting the dSph's formed a second generation of stars. 
This model probably will not work to explain UMi and Draco, because of the
age-metallicity relation and the star formation histories,
and seems to fail in general.

\acknowledgments
EO is grateful to Tad Pryor, Hans Rix, Taft Armandroff, and Kyle Cudworth.
My work is supported by the NSF through grant AST-9223967.

\end{document}